\pdfoutput=1
\documentclass[aps,floatfix,nofootinbib,onecolumn,superscriptaddress]{revtex4}
\usepackage{amsmath}
\usepackage{amssymb}
\usepackage{amsthm}
\usepackage{dcolumn}
\usepackage{epsfig}
\usepackage{graphics}
\usepackage{graphicx}
\usepackage{amsmath}
\usepackage{longtable}
\usepackage{color}

\definecolor{darkgreen}{rgb}{0,0.5,0}
\definecolor{purple}{rgb}{0.5,0,0.5}
\definecolor{nblue}{rgb}{0.0,0.0,0.50}
\definecolor{scarlet}{rgb}{1.0,0.2,0}
\usepackage[colorlinks=true, pdfstartview=FitV, linkcolor=purple, citecolor= purple, urlcolor=blue]{hyperref}

\begin{document}

{\par\raggedleft \texttt{SLAC-PUB-14487}\par}
{\par\raggedleft \texttt{CERN-LHeC-Note-2011-002 PHY}\par}
\bigskip{}

\title{Novel QCD Phenomenology at the LHeC}

\author{STANLEY J. BRODSKY$^*$ }

\affiliation {SLAC National Accelerator Laboratory\\
Stanford University, Stanford, California 94309 \\
and  CP$^3$-Origins, Southern Denmark University \\
Odense, Denmark\\
$^*$E-mail: sjbth@slac.stanford.edu\\}


\begin{abstract}

The proposed electron-proton/ion collider at CERN, the LHeC, can test fundamental and novel aspects of QCD and electroweak interactions as well as explore physics  beyond the standard model  over an exceptionally large kinematic range.

\end{abstract}

\maketitle

\vspace{10pt}

\section{Introduction}

The collisions of the LHC $7$ TeV proton beam with an electron beam with an energy of order $60$ GeV  would provide an extraordinary  facility for testing QCD, the electroweak interactions,  and theories beyond the standard model~\cite{Klein:2008zza}.   The $e p$ center of mass energy at the LHeC ranges up to
$\sqrt s = \simeq  1.3$ TeV,  about 30 times the CM energy range range of $e p$ collisions at HERA.

If new phenomena such as supersymmetry, the Higgs, heavy bosons, new quarks, etc, are discovered in $p p$ collisions at the LHC in the sub-TeV domain, the LHeC, with its smaller backgrounds, can be of great importance in illuminating the nature of the new physics.  Since quarks,  as well as electrons, radiate photons and the $W$ and $Z$ electroweak bosons, the LHeC can be considered a vector-boson collider similar in spirit to the proposed ILC $\gamma \gamma$ collider.  One can study in principle processes such as $W^+ W^- \to Z$ as well as  $\gamma \gamma \to H$, $Z Z \to H$, etc. The LHeC is especially well suited for the production of new heavy quarks $\tilde q$ in $e q^\prime \to e q^\prime $ collisions and leptoquark formation in $e q  \to L_q$ reactions.  It should be recalled that the H1 collaboration~\cite{Adloff:1997fg}
at HERA in fact observed an anomalous number of events in deep inelastic $e p$ charged- and neutral-current events at  $Q^2 > 15000$ GeV$^2$.

A primary test of QCD is deep inelastic electron-proton scattering $e p \to e^\prime X.$    If electron scattering at small angles can be detected, the minimum $x_{bj} = {Q^2 / 2 p{\cdot} q}$    with
$Q^2 \ge 1$ GeV$^2$ is $x_{min}  \simeq 10^{-6}$, about  one thousand times smaller than the HERA range.  In this very small $x$ domain one studies QCD at a high gluon density where saturation, BFKL, and other non-linear physics enter in an essential way.  Mueller~\cite{Mueller:1993rr} has shown that the power-law behavior of the gluon structure function at small $x$ $g(x,Q^2) \propto x^{1-\alpha}$ measures the intercept of the BFKL trajectory $\alpha = \alpha_{\rm BFKL}(t=0).$

The maximum momentum transfer of the electron-quark collision which can be studied at the LHeC ranges up to 0.9 TeV.   In this domain one tests the fundamental scaling of the neutral-current electron-quark interaction resulting from QCD evolution and $Z^0$ exchange at a distance scale of  order of $10^{-4} $ fm.  Similarly, by studying events with missing transverse energy, one can measure the primary weak interaction $e q \to \nu_e q$  due to $W$ exchange at a momentum transfer $Q$ as high as $100~ M_W.$  Events with single top and top pair can be studied.

The collisions of the LHC heavy ion beams with the $60$ GeV electron beam greatly increases the study of QCD and electroweak interactions in the nuclear domain, processes never measured at HERA.

\section{Novel QCD Topics}

\begin{enumerate}

\item {\it Diffractive Deep Inelastic Scattering  (DDIS) }

The final-state interactions of the struck quark with the spectators~\cite{Brodsky:2002ue} of the proton  in electron-quark collisions lead to diffractive events in deep inelastic scattering (DDIS) at leading twist,  such as $e p \to e^\prime p^\prime X ,$ where the proton remains intact and isolated in rapidity;    in fact, more than 10\% of the deep inelastic lepton-proton scattering events observed at HERA are
diffractive~\cite{Adloff:1997sc,Breitweg:1998gc}.  The presence of a rapidity gap
between the target and diffractive system requires that the target
remnant emerges in a color-singlet state; this is made possible in
any gauge by the soft rescattering incorporated in the Wilson line or by augmented light-front wavefunctions~\cite{Brodsky:2010vs}.  DDIS can also be studied in electroweak collisions, especially
$e p \to \nu_e p^\prime X ,$ where again the proton remains intact and isolated in rapidity.
The LHeC will test DDIS in extreme domains.  It can also be tested in nuclear collisions $e p \to e^\prime p^\prime X ,$ where the nucleus remains intact and isolated in rapidity.

\item{\it Deeply Virtual Compton Scattering}

One can study exclusive $e p$ reactions at high photon virtuality particularly deeply virtual Compton scattering $\gamma^* p \to \gamma p^\prime $  and its extensions, such as $\gamma^* p \to Z^0 p^\prime.$  The imaginary part of the DVCS amplitude in the handbag approximation determines the generalized parton distributions of the proton, and thus it is of particular interest.  The real part of the DVCS amplitude is determined by the  local  two-photon interactions of the quark current in the QCD light-front Hamiltonian~\cite{Brodsky:2008qu,Brodsky:1971zh}.  This contact interaction leads to a real energy-independent contribution to the DVCS amplitude  which is independent of the photon virtuality at fixed  $t$.  The interference of the timelike DVCS amplitude with the Bethe-Heitler amplitude leads to a charge asymmetry in $\gamma p \to \ell^+ \ell^- p$~\cite{Brodsky:1971zh,Brodsky:1973hm,Brodsky:1972vv}.    Such measurements can verify that quarks carry the fundamental electromagnetic current within hadrons.

\item{\it Deeply Virtual Exclusive Meson Production and Color Transparency}

Vector mesons and pseudoscalar mesons can be produced in hard deeply virtual meson scattering reactions: exclusive reactions such as $\gamma^* p \to V^0 p$ and $W^* p \to V^+ p.$    The distribution amplitude $\phi_M(x,Q^2)$ of the meson enters, a fundamental gauge-independent measure of the meson wavefunction.  In the case of electron-ion collisions one
can study color transparency~\cite{Brodsky:1988xz} at extreme conditions by checking the $A$ dependence of the quasielastic process  $\gamma^* A \to V^0 p X $.   Color transparency predicts that the cross section will be linear in the proton number $Z$ at high $Q^2$; i.e., no absorption of the produced vector meson as it traverses the nucleus since it is produced and emerges as a small color singlet.

\item{\it Color-Transparent Higher-Twist Direct Processes}

It is conventional to assume that high transverse momentum hadrons in inclusive reactions arise only from jet fragmentation. In fact high $p^H_\perp$ events  can emerge
directly from a hard higher-twist subprocess~\cite{Arleo:2009ch,Arleo:2010yg,Arleo:2010zz}.
This phenomena can explain~\cite{Brodsky:2008qp} the  ``baryon anomaly" observed at RHIC --  the ratio of baryons to mesons at high $p^H_\perp$,  as well as the power-law fall-off $1/ p_\perp^n$ at fixed $x_\perp = 2 p^H_\perp/\sqrt s, $ both  increase with centrality~\cite{Adler:2003kg}, opposite to the usual expectation that protons should suffer more energy loss in the nuclear medium than mesons.

A  fundamental test of leading-twist QCD predictions in high transverse momentum reactions is the measurement of the power-law
fall-off of the inclusive cross section~\cite{Sivers:1975dg}
${E d \sigma/d^3p}(A B \to C X) ={ F(\theta_{cm}, x_T)/ p_T^{n_eff} } $ at fixed $x_T = 2 p_T/\sqrt s$
and fixed $\theta_{CM},$ where $n_{eff} \sim 4 + \delta$. Here $\delta  =  {\cal O}(1)$ is the correction to the conformal prediction arising
from the QCD running coupling and the DGLAP evolution of the input distribution and fragmentation functions~\cite{Brodsky:2005fz,Arleo:2009ch,Arleo:2010yg}.
The direct higher-twist subprocesses, where the trigger hadron is produced  within the hard subprocess avoid the waste of same-side energy, thus allowing the target  structure functions to be evaluated at the minimum values of parton momenta where they are at their maximum.

In the case of electron-proton collisions, one can measure $e p \to e H X$ or $\gamma p \to H X$ processes where the hadron $H$ emerges in isolation.   This reaction is particularly interesting to study in electron-ion collisions at the LHeC since the hadron is initially produced as a small-size $b_\perp \sim 1/p_T$  color-singlet state; it is ``color transparent" ~\cite{Brodsky:1988xz},   and it can thus propagate through dense nuclear matter with minimal energy loss.  In
contrast, the hadrons which are  produced from jet fragmentation have a  normal inelastic cross section.

The power law of the cross section $E d \sigma/d^3p ( e p \to e H X)$ at fixed $x_T = p^H_\perp/\sqrt s$ and $\theta_{cm}$ can differ from leading-twist predictions due to the presence of higher-twist processes.
The behavior of the single-particle inclusive cross section is thus a key test of QCD at the LHeC, since the leading-twist prediction for $n_{\rm eff} \sim 4 + \delta$ is independent of the detailed form of the structure and fragmentation functions.

\item {\it Sivers and Collins Effects}

The effects of final-state interactions of the scattered quark  in deep inelastic scattering  have been traditionally assumed to be power-law suppressed.  In fact,  the final-state gluonic interactions of the scattered quark lead to a  $T$-odd non-zero spin correlation of the plane of the lepton-quark scattering plane with the polarization of the target proton~\cite{Brodsky:2002cx}.  This  leading-twist Bjorken-scaling ``Sivers effect"  is nonuniversal since QCD predicts an opposite-sign correlation~\cite{Collins:2002kn,Brodsky:2002rv} in Drell-Yan reactions due to the initial-state interactions of the annihilating antiquark.   Similarly, the final-state interactions of the produced quark with its comoving spectators produces a final-state $T$-odd polarization correlation -- the ``Collins effect.".  This can be measured at the LHeC without beam polarization by measuring the correlation of the polarization of a hadron such as the $\Lambda$ with the quark-jet production plane.

\newpage
\item{\it The Odderon}

QCD predicts an odd-$C$ exchange trajectory due to diagrams with three-gluon exchange at lowest order, a fundamental effect which has never been verified. The BFKL analysis predicts that its trajectory has an intercept
$\alpha_{Odderon}(t=0) \simeq 0$. The odderon can be measured in processes requiring odd-$C$ exchange such as $\gamma p \to \pi^0 p^\prime.$  An even more sensitive test, ideal for the LHeC is to measure the difference between the charm and anti-charm angular or energy distributions in $\gamma^* p \to c \bar c p^\prime$~\cite{Brodsky:1999mz}. The asymmetry arises from the interference of the pomeron and odderon exchange amplitudes.

\item{\it Non-Universal Antishadowing}

The nuclear distribution can be measured to extraordinarily small $x_{Bj}$ in electron-ion collisions at the LHeC.

The shadowing of the nuclear structure functions: $R_A(x,Q^2) < 1 $ at small $x < 0.1 $ can be readily understood in terms of the Gribov-Glauber
theory.  Consider a two-step process in the nuclear target rest frame. The incoming $q \bar q$ dipole first interacts diffractively $\gamma^*
N_1 \to (q \bar q) N_1$ on nucleon $N_1$ leaving it intact.  This is the leading-twist diffractive deep inelastic scattering  (DDIS) process
which has been measured at HERA to constitute approximately 10\% of the DIS cross section at high energies.  The $q \bar q$ state then interacts
inelastically on a downstream nucleon $N_2:$ $(q \bar q) N_2 \to X$. The phase of the pomeron-dominated DDIS amplitude is close to imaginary,
and the Glauber cut provides another phase $i$, so that the two-step process has opposite  phase and  destructively interferes with the one-step
DIS process $\gamma* N_2 \to X$ where $N_1$ acts as an unscattered spectator. The one-step and-two-step amplitudes can coherently interfere as
long as the momentum transfer to the nucleon $N_1$ is sufficiently small that it remains in the nuclear target;  {\em i.e.}, the Ioffe
length~\cite{Ioffe:1969kf} $L_I = { 2 M \nu/ Q^2} $ is large compared to the inter-nucleon separation. In effect, the flux reaching the interior
nucleons is diminished, thus reducing the number of effective nucleons and $R_A(x,Q^2) < 1.$
The Bjorken-scaling diffractive contribution to DIS arises from the rescattering of the struck quark after it is struck  (in the
parton model frame $q^+ \le 0$), an effect induced by the Wilson line connecting the currents. Thus one cannot attribute DDIS to the physics of
the target nucleon computed in isolation~\cite{Brodsky:2002ue}.

Nuclear shadowing and antishadowing in electron-ion collisions  is  thus due to multi-step coherent reactions involving
leading twist diffractive reactions~\cite{Brodsky:1989qz,Brodsky:2004qa}. The nuclear shadowing of structure functions is a consequence of
the lepton-nucleus collision; it is not an intrinsic property of the nuclear wavefunction.

The {\it antishadowing} of the nuclear structure functions as observed in deep
inelastic lepton-nucleus scattering is particularly interesting. Empirically, one finds $R_A(x,Q^2) \equiv  \left(F_{2A}(x,Q^2)/ (A/2) F_{d}(x,Q^2)\right)
> 1 $ in the domain $0.1 < x < 0.2$;  {\em i.e.}, the measured nuclear structure function (referenced to the deuteron) is larger than the
scattering on a set of $A$ independent nucleons.

Ivan Schmidt, Jian-Jun Yang, and I~\cite{Brodsky:2004qa} have extended the analysis of nuclear shadowing  to the shadowing and antishadowing of the
electroweak structure functions.  We note that there are leading-twist diffractive contributions $\gamma^* N_1 \to (q \bar q) N_1$  arising from Reggeon exchanges in the
$t$-channel~\cite{Brodsky:1989qz}.  For example, isospin--non-singlet $C=+$ Reggeons contribute to the difference of proton and neutron
structure functions, giving the characteristic Kuti-Weisskopf $F_{2p} - F_{2n} \sim x^{1-\alpha_R(0)} \sim x^{0.5}$ behavior at small $x$. The
$x$ dependence of the structure functions reflects the Regge behavior $\nu^{\alpha_R(0)} $ of the virtual Compton amplitude at fixed $Q^2$ and
$t=0.$ The phase of the diffractive amplitude is determined by analyticity and crossing to be proportional to $-1+ i$ for $\alpha_R=0.5,$ which
together with the phase from the Glauber cut, leads to {\it constructive} interference of the diffractive and nondiffractive multi-step nuclear
amplitudes.  The nuclear structure function is predicted to be enhanced precisely in the domain $0.1 < x <0.2$ where
antishadowing is empirically observed.  The strength of the Reggeon amplitudes is fixed by the fits to the nucleon structure functions, so there
is little model dependence.
Since quarks of different flavors  will couple to different Reggeons; this leads to the remarkable prediction that
nuclear antishadowing is not universal; it depends on the quantum numbers of the struck quark. This picture implies substantially different
antishadowing for charged and neutral current reactions, thus affecting the extraction of the weak-mixing angle $\theta_W$.  The ratio of nuclear to nucleon structure functions $R_{A/N}(x,Q) = {F_{2A}(x,Q)\over A F_{2N}(x,Q)}$ is thus process independent.   We have also identified
contributions to the nuclear multi-step reactions which arise from odderon exchange and hidden color degrees of freedom in the nuclear
wavefunction.

\begin{figure}[!]
 \begin{center}
\includegraphics[width=18.0cm]{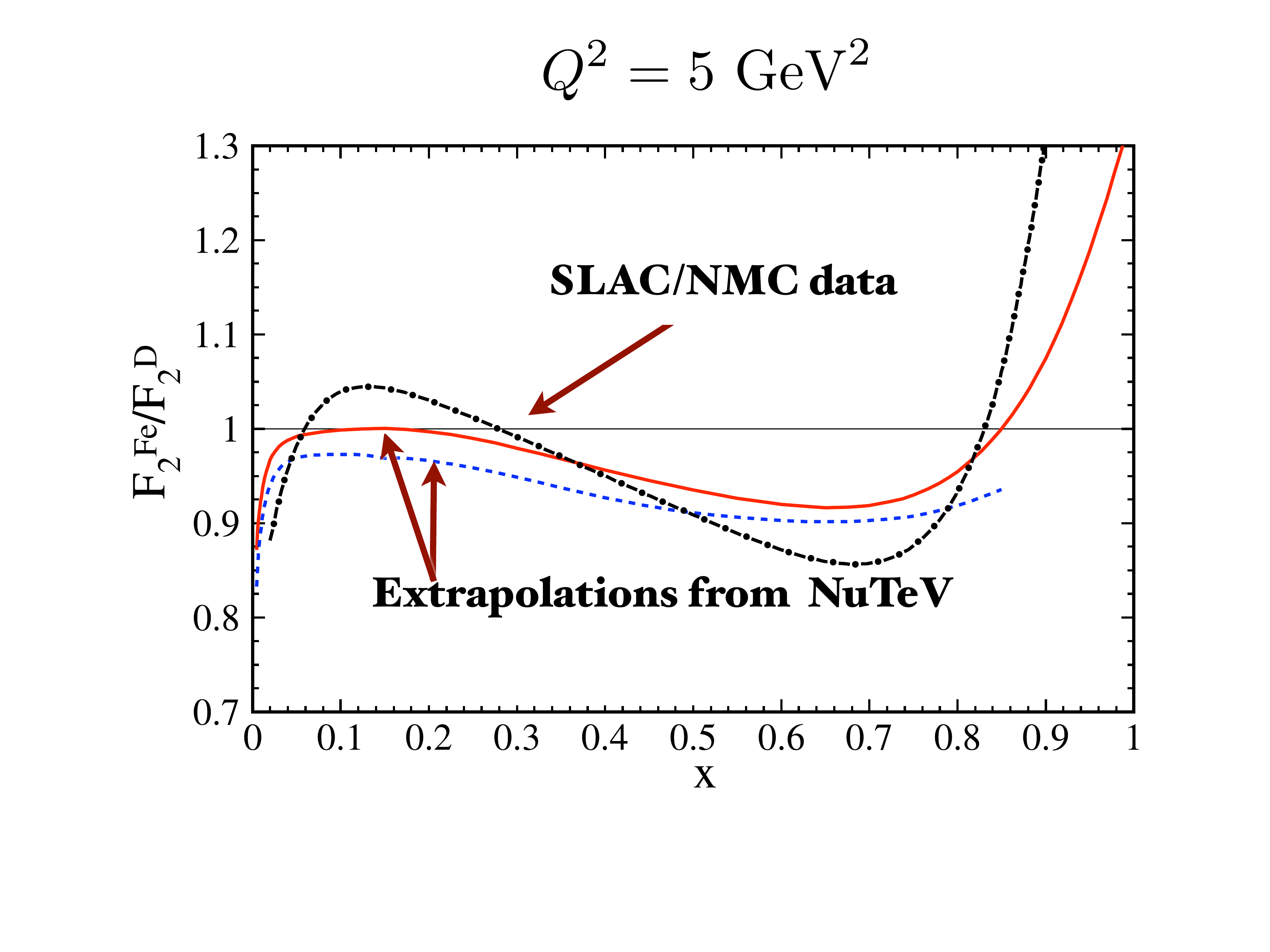}
\end{center}
  \caption{Comparison of the Nuclear Modification  of Charged vs. Neutral Current Deep Inelastic Structure Functions.  From  I.~Schienbein et al. ~\cite{Schienbein:2008ay} }
\label{figNew9}
\end{figure}

It has been conventional  to assume that the nuclear modifications to the structure functions measured in deep inelastic lepton-nucleus and neutrino-nucleus interactions are identical;  in fact,  as noted above, the Gribov-Glauber theory predicts that the antishadowing of nuclear structure functions in the $x \sim 0.15$ domain  is not  universal, but depends on the quantum numbers of each struck quark and antiquark~\cite{Brodsky:2004qa} because of flavor-dependent Regge exchange.  This observation can explain the recent analyses  of Schienbein et al.\cite{Schienbein:2008ay}  and Kovarik et al.  \cite{Kovarik:2010uv}  which shows that the measured nuclear effect measured in the NuTeV deep inelastic scattering charged current experiment  is distinctly different from the nuclear modification measured at SLAC and NMC in deep inelastic scattering electron and muon scattering.   See fig.\ref{figNew9}.    This implies that part of
of the anomalous NuTeV result~\cite{Zeller:2001hh} for $\theta_W$ could be due to the non-universality of nuclear antishadowing for charged and
neutral currents.

This nonuniversality can be tested at the LHeC in electron-ion collisions by comparing the $A$-dependence of deep inelastic neutral and charged current reactions in the $x \sim 0.15$ domain or by  tagging the jet flavor in semi-inclusive DIS.

\item{\it Hidden Color}

In nuclear physics nuclei are simply the composites of nucleons. However, QCD provides a new perspective:~\cite{Brodsky:1976rz,Matveev:1977xt}  six quarks in the fundamental
$3_C$ representation of $SU(3)$ color can combine into five different color-singlet combinations, only one of which corresponds to a proton and
neutron.  The deuteron wavefunction is a proton-neutron bound state at large distances, but as the quark separation becomes smaller,
QCD evolution due to gluon exchange introduces four other ``hidden color" states into the deuteron
wavefunction~\cite{Brodsky:1983vf}.  The normalization of the deuteron form factor observed at large $Q^2$~\cite{Arnold:1975dd}, as well as the
presence of two mass scales in the scaling behavior of the reduced deuteron form factor~\cite{Brodsky:1976rz}, suggest sizable hidden-color
Fock state contributions  in the deuteron
wavefunction~\cite{Farrar:1991qi}.
The hidden-color states of the deuteron can be materialized at the hadron level as   $\Delta^{++}(uuu)\Delta^{-}(ddd)$ and other novel quantum
fluctuations of the deuteron. These dual hadronic components become important as one probes the deuteron at short distances, such
as in exclusive reactions at large momentum transfer.  For example, the ratio  ${{d \sigma/ dt}(\gamma d \to \Delta^{++}
\Delta^{-})/{d\sigma/dt}(\gamma d\to n p) }$ is predicted to increase to  a fixed ratio $2:5$ with increasing transverse momentum $p_T.$
Similarly, the Coulomb dissociation of the deuteron into various exclusive channels
$e d \to e^\prime + p n, p p \pi^-,  \Delta \Delta, \cdots $
will have a changing composition as the final-state hadrons are probed at high transverse momentum, reflecting the onset of hidden-color
degrees of freedom.
The hidden color of the deuteron can be probed at the LHeC in electron deuteron collisions by studying reactions such as $\gamma^* d \to n p X$ where the proton and neutron emerge in the target fragmentation region at high and opposite $p_T$.   In principle, one can also study DIS reactions $e d \to e^\prime X$ at very high $Q^2$ where $x > 1.$  The production of high $p_T$ anti-nuclei at the LHeC is also sensitive to hidden color-nuclear components.

\item	 {\it Heavy Quark Distributions}

It is conventional to assume that the charm and bottom quarks in the proton structure function  only arise from gluon splitting $g \to Q \bar Q.$  In fact, the proton light-front wavefunction contains {\it ab initio } intrinsic heavy quark Fock state components such as $|uud c \bar c>$~\cite{Brodsky:1980pb,Brodsky:1984nx,Harris:1995jx,Franz:2000ee}.   Intrinsic heavy quark Fock states in the hadron light-front wavefunction are a rigorous consequence of QCD. The intrinsic heavy quarks carry most of the proton's momentum since this minimizes the off-shellness of the state. The heavy quark pair $Q \bar Q$ in the intrinsic Fock state  is primarily a color-octet,  and the ratio of intrinsic charm to intrinsic bottom scales scales as $m_c^2/m_b^2 \simeq 1/10,$ as can be seen from the operator product expansion in non-Abelian QCD~\cite{Brodsky:1984nx,Franz:2000ee}.   Intrinsic charm and bottom explain the origin of high $x_F$ open-charm and open-bottom hadron production, double charm baryon production, as well as the single and double $J/\psi$ hadroproduction cross sections observed at high $x_F$.   The factorization-breaking nuclear $A^\alpha(x_F)$ dependence  of hadronic $J/\psi$ production cross sections can also be explained~\cite{Brodsky:2006wb}.

As emphasized by Lai, Tung, and Pumplin~\cite{Pumplin:2007wg}, there are strong indications that the structure functions used to model charm
and bottom quarks in the proton at large $x_{bj}$ have been underestimated, since they ignore intrinsic heavy quark fluctuations of
hadron wavefunctions.
Furthermore, the neglect of the intrinsic-heavy quark component in the proton structure function will lead to an incorrect assessment of the gluon distribution at large $x$ if it is assumed that sea quarks always arise from gluon splitting~\cite{Stavreva:2010mw}
Intrinsic heavy quarks in the  $|uud q \bar q>$ Fock states can  account for the magnitude and shape of the
proton's
 $\bar d - \bar u$, $s + \bar s$, and $\bar u + \bar d - s -\bar s$ distributions~\cite{Chang:2011du}.

\begin{figure}[!]
 \begin{center}
\includegraphics[width=16.0cm]{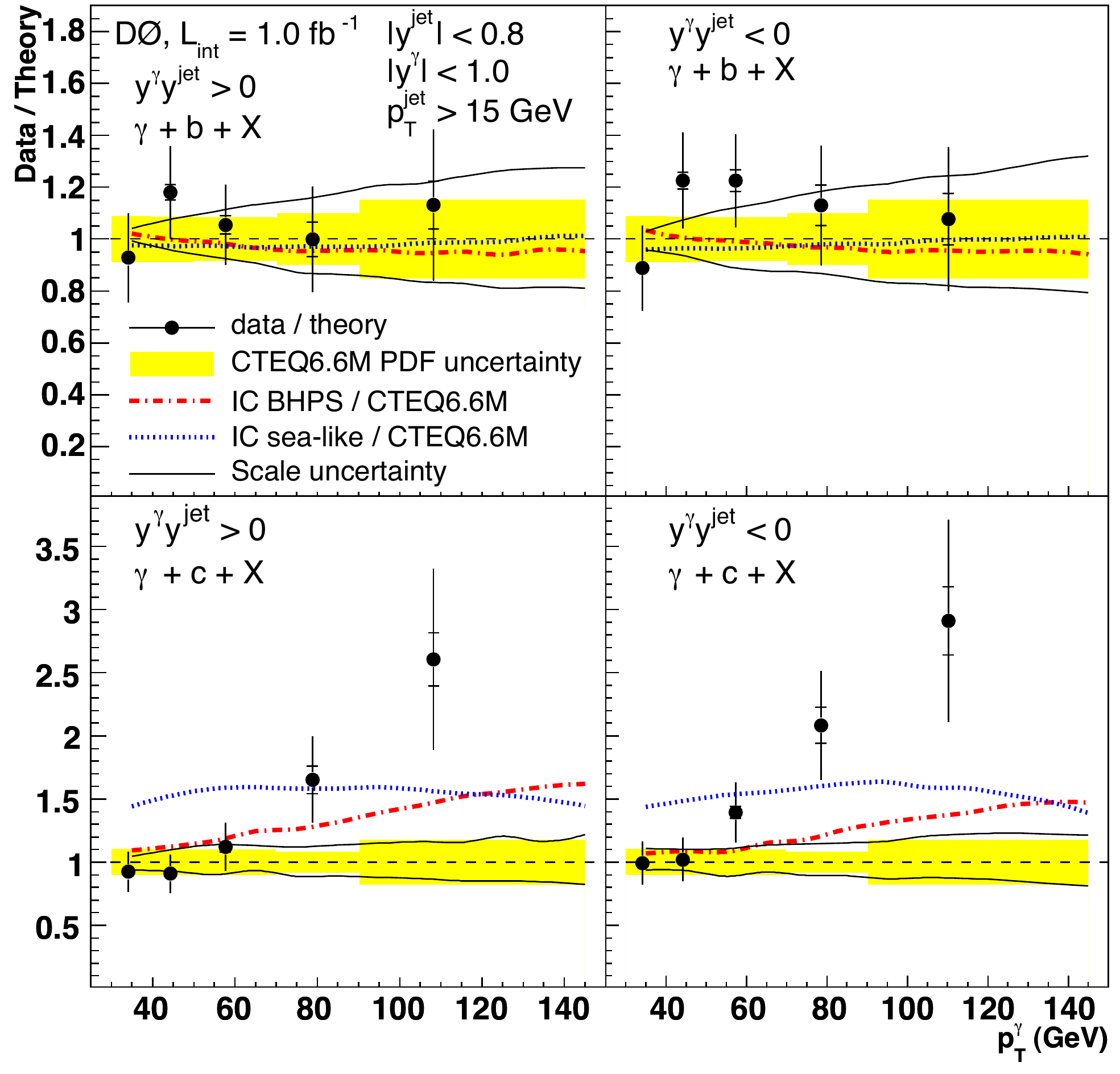}
\end{center}
  \caption{Measurement of the $p \bar p \to \gamma  b X$ and $p \bar p \to \gamma  c X$ cross sections from Dzero.  From ref.  ~\cite{Abazov:2009de} }
\label{figDzero}
\end{figure}

The anomalous growth of the $p \bar p \to \gamma  c X$ inclusive cross section observed by D0 collaboration~\cite{Abazov:2009de} at the Tevatron indicates that the charm  distribution has been underestimated at $x> 0.10.$
See fig. \ref{figDzero}.
The LHeC could definitively establish the phenomenology of the charm  and bottom structure functions at
large $x_{bj}.$    In addition to DIS measurements, one can test the charm (and bottom) distributions at the LHeC  by measuring reactions such as $\gamma p \to c X$ where the charm jet is produced at high $p_T$ in the reaction $\gamma c \to c g$.

Kopeliovich, Schmidt, Soffer, Goldhaber, and I ~\cite{Brodsky:2006wb} have  proposed a novel mechanism for Inclusive and diffractive
Higgs production $pp \to p H p $ in which the Higgs boson carries a significant fraction of the projectile proton momentum. The production
mechanism is based on the subprocess $(Q \bar Q) g \to H $ where the $Q \bar Q$ in the $|uud Q \bar Q>$ intrinsic heavy quark Fock state of the colliding proton has approximately
$80\%$ of the projectile protons momentum.   A similar mechanism could produce the Higgs at large $x_F \sim 0.8$ in $\gamma p \to H X$ at the LHeC based on the mechanism
$\gamma (Q \bar Q)  \to H $ since the heavy quarks typically each carry light-front momentum fractions $x \sim 0.4$ when they arise from the intrinsic heavy quark Fock states $|uud Q \bar Q>$ of the proton.

\item	 {\it Top Quark Production-- Anomalous Threshold Effects}

The top quark can be produced the LHeC from $W b \to t$ reactions where the $b$ quark arises from the intrinsic bottom component. One can also produce top-quark pairs in $\gamma^* p \to t \bar t X$ reactions.
Since the top quarks are produced non-relativistically, there are important QCD corrections~\cite{Brodsky:1995ds} which increase their production near threshold -- the analog of the QED Sakharov, Sommerfeld, Schwinger corrections. This effect involves the QCD coupling $\alpha_s(\mu^2)$ at a renormalization scale of order $\mu^2 \propto \beta^2 \hat s$, where $\beta$ is the $Q \bar Q$ relative velocity and $\hat s = 4 m^2_T.$

\item{\it Renormalization Scales at the LHeC} 	

It is often stated that the renormalization scale of the QCD running coupling $\alpha_s(\mu^2_R) $  cannot be fixed, and thus it has to be chosen in an {\it ad hoc} fashion.  In fact, as in QED, the scale can be fixed unambiguously by shifting $\mu_R$  so that all terms associated with the QCD $\beta$ function vanish.  The purpose of the running coupling in any gauge theory is to sum all
terms involving the $\beta$ function; in fact, when the
renormalization scale is set properly, all non-conformal
$\beta \ne 0$ terms in a perturbative expansion arising from
renormalization are summed into the running coupling. In general, each set of skeleton diagrams has its respective scale.  The remaining terms in the perturbative series are then identical to that of a conformal theory; i.e., the corresponding theory with $\beta=0$.  The resulting scale-fixed predictions using this  ``principle of maximum conformality" (PMC) ~\cite{Brodsky:2011zz} are independent of the choice of
renormalization scheme --  a key requirement of renormalization
group invariance.  The result is independent of the choice of the initial renormalization scale ${\mu_R}_0$, thus satisfying Callan-Symanzik invariance.   The results
agree with QED scale-setting in the Abelian limit. This is also the theoretical principle
underlying the BLM procedure~\cite{Brodsky:1982gc} , commensurate scale relations between observables, and the scale-setting method used in lattice gauge theory.  The number of active flavors $n_f$ in the QCD $\beta$ function is also correctly determined.

Unlike heuristic scale-setting procedures, the BLM/PMC method~\cite{Brodsky:1982gc,Brodsky:2011zz} gives results which are independent of the choice of renormalization scheme, as required by the transitivity property of the renormalization group.
The divergent renormalon terms of order $\alpha_s^n \beta^n n!$ are transferred to the physics of the running coupling.                                                                      Furthermore, one retains sensitivity to ``conformal' effects which arise in higher orders, physical effects which are not associated with QCD  renormalization.  The BLM/PMC  method also provides scale-fixed,
scheme-independent high precision connections between observables, such as the ``Generalized Crewther Relation" ~\cite{Brodsky:1995tb}, as well as other ``Commensurate Scale Relations" ~\cite{Brodsky:1994eh,Brodsky:2000cr}.   In the case of  the three-gluon coupling, the renormalization scale has a unique dependence on the virtualities of the three gluons~\cite{Binger:2006sj}.

The elimination of the renormalization
scheme ambiguity using the BLM/PMC method will not only increase the
precision of QCD tests,  but it will also increase the sensitivity
of LHeC experiments to new physics beyond the Standard Model.

\end{enumerate}

\newpage

\vspace{15pt}

\noindent{\bf Acknowledgments}

\vspace{5pt}

I thank Max Klein for motivating this report and for his remarkable leadership of the LHeC project.
This research was supported by the Department
of Energy,  contract DE--AC02--76SF00515.

\end{document}